# Human Mars Mission Architecture Plan to Settle the Red Planet with 1000 People

Malaya Kumar Biswal M[1], Vishnu S[2], Devika S Kumar[3], Sairam M[4]

Pondicherry University, Kalapet, Puducherry, India - 605 014

**Abstract**

Exploration is one of the attentive endeavor to mankind and a strategy for evolution. We have been incessantly reconnoitering our planet and universe from Mesopotamian era to modern era. The progression of rocketry and planetary science in past century engendered a futuristic window to explore Mars which have been a source of inspiration to hundreds of astronomers and scientists. Globally, it invigorated space exploration agencies to make expedition for planetary exploration to Mars and Human Mars Missions. Scientists and engineers have portrayed numerous Human Mars Mission proposals and plans but currently the design reference mission 5.0 of NASA is the only mission under study. Here we propose a mission architecture for permanent Human Mars Settlement with 1000 peoples with multiple launch of sufficient cargoes and scientific instruments.

**Introduction:**

This paper focuses on design of Human Mars Mission with reference to the instructions by Mars Society. We proposed mission architecture for carrying 1000 peoples onboard spaceship (Marship). Overall mission architecture outline map and Human Mars Settlement Map is provided next to this page. We divided the whole mission architecture into three phases starting from orbital launch of launch vehicles and Mars colony establishment. We proposed novel habitat for protection during robust dust storms, various method to make the colony economically successful, minerals and their applications, administrative methods, water extraction, plantation, landing patterns, estimation of masses of food to be carried out and customizable system for re-use and recycling. Further, we suggested list of plants required for faster food production and oxygen production. The whole colony will be in an ordered pattern shown next to this page called "Human Mars Settlement Map".

**Technology Considerations in this paper:**

| | |
|---|---|
| • High scale payload delivery launch vehicle<br>• Dual Communication System Fig.2<br>• Deployable Solar Panel with dust-proof<br>• Mars Sub-surface Habitat<br>• Mars Sub-surface drillers for water<br>• Auto Cropping System<br>• Auto Atmosphere Regulation System<br>• Auto Soil Fertilizing Rover System | • Solar Cell Radiation Shields<br>• Foldable Solar Power Plant<br>• Conveyor Belt Roving Rovers<br>• International Mars Station<br>• Auto Trajectory Correction maneuver<br>• Auto Scientific landing detection<br>• Fast growing plants for food production<br>• Fast Oxygen producing plants<br>• In-Situ Resource Utilization |

---

[1] Department of Physics, Pondicherry University, Kalapet, Puducherry, India – 605014. Contact:malaykumar1997@gmail.com

[2] Department of Computer Science, Sri Manakula Vinayagar Engineering College, India. Contact: vishnucse57@gmail.com
[3] Department of Earth Sciences, Pondicherry University, Kalapet, Puducherry, India – 605 014. Contact: devikask412@gmail.com
[4] Department of Mechanical Engineering, Sri Venkateswara College of Engineering, Puducherry, India. Contact: sairamsiva198@gmail.com

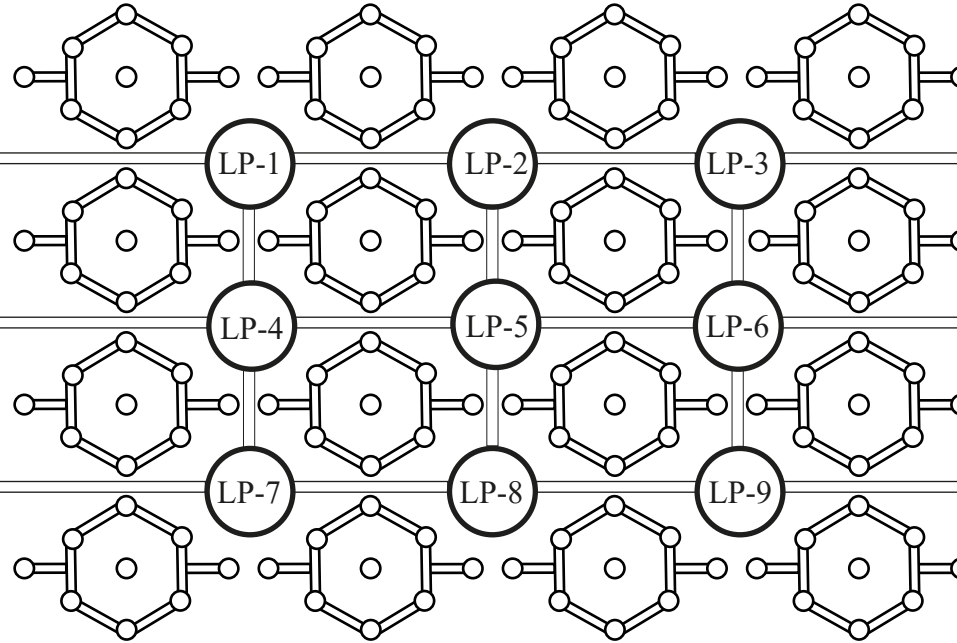

| | | |
|---|---|---|
| EXPORT LAUNCH PAD - I | ADMINISTRATIVE BUILDING | SURVEILLANCE PILLAR - I |
| SHOP - I | -- PLATFORM FOR MOVEMENT OF LOAD / PASSENGER ROVERS -- | IRON PRODUCTION PLANT |
| MARS MINERALS EXTRACTION & FABRICATION PLANT | LANDING OF MODULES IN A HEXAGONAL PATTERN | WATER EXTRACTION & OXYGEN PRODUCTION PLANT |
| SOLAR CELL PRODUCTION PLANT | WAY - I ... WAY - II | MARS PROPELLANT PLANT |
| ROVER MECHANICAL SYSTEM REPAIRING SHED | MARS COLONY REPLICATION - WITH 16 BASES | 3D PRINTING PLANT |
| SHOP - II | -- PLATFORM FOR MOVEMENT OF LOAD / PASSENGER ROVERS -- | ARTIFICIAL INTELLINGENCE ROBOT ASSEMBLY PLANT |
| EXPORT LAUNCH PAD - II | HOSPITAL   GLASS HOUSE MARS SURFACE GARDEN   RESTAURANT | SURVEILLANCE PILLAR - II |

Landing Pads: LP-1, LP-2, LP-3, LP-4, LP-5, LP-6, LP-7, LP-8, LP-9

```
                        ┌──────────────────────────────────┐
                        │ Human Mars Mission Architecture  │
                        └──────────────────────────────────┘
                                         │
            ┌────────────────────────────┼────────────────────────────┐
            ▼                            ▼                            ▼
       ┌─────────┐                  ┌──────────┐                 ┌───────────┐
       │ PHASE-I │                  │ PHASE-II │                 │ PHASE-III │
       └─────────┘                  └──────────┘                 └───────────┘
```

- **PHASE-I**
  - **Launch**
    - Cost and Capability
      - Big Falcon rocket
    - Payload Capability
      - 1 to 2 tons to Mars
    - Cost Per Launch
      - $5 Billion USD
  - **Spaceship Rendezvous**
    - Marship-I
      - Ascent Vehicle-I
      - Medical Module
      - Life Support System Module
      - Crewed Module
      - Crop Culture Module
      - Cargo Module
      - Fuel Tank & Thrusters
    - Marship-II
      - Ascent Vehicle-I
      - Habitat-I
      - Habitat-II
      - Rover Module
      - Lander Module
      - Cargo Module
      - Fuel Tank & Thrusters
  - **Earth-Mars Departure**
    - Sufficient Thrust
    - Hohmann Transfer Trajectory
    - Auto-pilot System
    - Auto Trajectory Correction
  - **Mars Approach**

- **PHASE-II**
  - **Preparation**
    - Ascent Vehicle I & II Separation
    - Medical Module Separation
    - Life Support System Separation
    - Crewed Module Separation
    - Crop Culture Module Separation
    - Cargo Module I & II Separation
    - Habitat I & II Separation
    - Rover Module Separation
    - Lander Module Separation
    - Empty Fuel tank direction
  - **Aerocapture**
    - Dual Layered Heat Shield
    - Hypersonic Inflatable Aerodynamic Decelerators
    - Retropropulsive Powered Propulsion
    - Parachute Deployment System
  - **Landing**
    - All Landing Modules
    - Good Impact Absorbers
    - Auto Landing Site Detection
    - Mapped Landing
  - **Surface Operation**
    - Executing and planning for Surface Exploration

- **PHASE-III**
  - **Setup & Synchronization**
    - Power Production Plant
    - Water Extraction Plant
    - Crop Culture Plant
    - Oxygen Production Plant
    - Linking all modules
    - Preparation for a Scientific Goal
  - **Surface Exploration**
    - Soil & Rock Sample Collection
    - Atmosphere Monitoring
    - Resource Location
    - Weather Monitoring
    - Asteroid Monitoring
    - Monitoring Dust Storms
    - Investigating Existence of Life
    - Analyzing Past Landed Missions
    - Doing Life Experiments
    - Adapting Life to the Martian Environment
  - **In-Situ Resource Utilization**
    - Propellant Production from $CO_2$
    - Water Extraction
    - Utilizing Rocks for Constructions
    - Utilizing Mars Minerals
    - Extraction of Ferrous
  - **Mars Colony Replication**
  - **Service & Economy**
  - **Administration**

# PHASE-I

**Launch:** Our overall Mars mission is classified into four phases (Phase-I, II, II & IV) where Phase IV is optional and opted for Return trip to Earth. In modern times human mission to Mars requires launch of enormous number of cargoes, crewed modules and scientific laboratories. Launch vehicle should have large payload delivery capability to Mars. As per current emerging technology, we prefer to use Spacex's Big Falcon Rocket among two major heavy launch vehicles (Falcon Heavy Expandable and Delta IV Heavy). Big Falcon Rocket (BFR) has 1 to 2 tons (100,000 kg to 200,000 kg) payload delivery capability to Mars. Hence utilizing BFR we wish to deliver all the mission modules into LEO (Low-earth Orbit) in two attempts. [1]

**Attempt-I:** In attempt-1 (2-3 BFR launches) BFR will be launched carrying Cargo module, fuel tanks, crewed module, life support system module, medical module and Mars ascent vehicle-I for Marship-I Rendezvous.

**Attempt-II:** In attempt-2 (2-3 BFR launches) BFR will be delivering cargo module, Mars landers, Mars rovers, Habitat-I & II and Mars ascent vehicle-II for Marship-II Rendezvous.

**Spaceship Rendezvous**: All the launched modules will be integrated and assembled in low-earth orbit to form Marship-I and Marship-II. Some of the features of spaceship were have artificial gravity system to simulate artificial gravity, good life support system, entertaining block, better radiation shield, air regulation system, effective pressurization system, recycling system of carbon dioxide and human waste, good electronic components and possess sufficient space for cargo storage. **Special operating features**: auto-repair system, asteroid detection, radiation zone detection, auto trajectory estimation and auto-trajectory maneuver correction.

**Earth-Mars Departure:** Subsequent to Spaceship Rendezvous, the Marship propels towards Mars with significant thrust over Hohmann transfer trajectory. Electrical components and other onboard spaceship components is powered by solar power generated by solar cells. The spaceship is capable of generating artificial gravity using AG-thrusters.

**Mars Approach:** Calculation using Hohmann Transfer Trajectory estimated that a Mars transit takes around 270-300 days (travel time). After a cruise travel the spaceship finally approaches Mars. [2]

# PHASE-II

**Preparation:** Following Mars approach, all the modules of Marship I & II get separated as shown in Fig.3 Before proceeding for atmospheric entry, all the modules and their communication systems were tested in advance and their communication links were interlinked with either International Mars Station or Mars Master Control Station for better communication and effective landing operations. Following this, all the modules should prefer orbital entry (i.e. entry from Mars orbit instead of direct entry [3]), and landed as per the Human Mars Settlement Map. All the lander have a common special feature of auto-detection of scientific landing site.

**Aerocapture:** After performing successful atmospheric entry, all the module is exposed to atmospheric heating due to sufficient thickness of Martian atmosphere. Hence we prefer to use dual layered heat shield (i.e. normal heat shield having extra water layered coat). This may effectuate good heat resistant and protection. Moreover all the lander modules fitted with

Hypersonic Inflatable Aerodynamic Decelerator (HIAD) for faster Aerocapture. Additionally, these landing modules also equipped with large sized parachute deployment system as well as Retropropulsive powered descent system for promoting effective soft-landing. (In consideration for manned or crewed landing).

**Landing:** Posterior to Aerocapture, landing occurs. All landing modules good impact absorption system like hydraulic impact attenuating systems. Landers possess auto-detection of scientific landing site and lands as per pattern showed in Human Mars Settlement Map.

**Surface Operation:** Cosmonauts from all the modules undergo for a discussion in perspective of present and future surface exploration. Here peoples makes decision what to be done on the planetary surface.

<p align="center">**PHASE-III**</p>

**Setup & Synchronization**: Modules were interlinked and power plant, communication system with IMS or MMCS is synchronized. Setting-up of Solar Power Plant, Water Extraction Plant, Crop Production Plant, In-Situ Propellant Plant, $CO_2$ and $O_2$ Production Plant and Surveillance Copters. Finally Initialization of scientific goals and mission as planned.

**Surface Exploration**:

- Analyzing the collection of Rock and Soil Samples
- Weather Monitoring
- Resource Location
- Atmospheric Changes Monitoring
- Asteroid Tracking (Counts and their impact sites)
- Monitoring Robust Dust Storms
- Investigation of existence of life
- Analyzing the results of past dead missions from their landing sites
- Doing life experiments
- Adapting life forms to accordance to the Martian environment.
- Training life forms to adapt to Martian gravity and atmosphere.

**In-Situ Resource Utilization:**

- Production of Mars propellant on Mars
- Extraction of Water from Sub-surface of Mars
- Utilizing Rocks and other minerals for Construction purpose
- Extracting, fabricating and utilizing Mars Minerals
- Utilizing and re-using scientific instruments from past dead missions

**Mars Colony Replication:**

Artificial Intelligence robots and assembling gadgets will start making subsequent base for future Martian generation. Mars colony replication is shown in Human Mars Settlement Map.

**Everyday Service:**

- Every Martian should aim for the prosperous of Martian community in both economic and technology.
- Every Martian should do scientific research, explore and reveal the secret behind life, explore the infinity beyond space. (Excludes the personal time of Martian like playing, entertaining etc…)
- Every Martian should do cooperative work towards the development of Mars colony.
- Everyday checkup of scientific instruments and production plants.
- Few group should always monitor the storms and asteroid impacts
- Every Martian should do exercise in order to make them healthy

**Making the Colony Economically Successful:**

- Selling the rock and soil samples to the technologically developing nations ( Nations that not reached Mars)
- Commissioning Service Missions (i.e. repairing earlier dead missions, dead rovers and landers) in sake of money.
- Providing Hospitality Services to the astronauts landed on Mars from earth (for the purpose of research or tourism) in sake of money.
- Making "Mars" a platform for Deep Space Network by building antennas and providing network relay from Earth and spacecrafts/spaceships beyond Mars in exchange of money.
- Making available of "Emergency Service Mission" during the critical stage of Mars transit between Earth-Mars and beyond for money.
- Guiding the spacecrafts and landers via direct control signals from earth may cause interruption and there may be chances of losing the entire missions. Hence all the trajectory and EDL guidance will be undertaken by Mars Master Control Station in Mars orbit for money.
- In future, refueling of spaceships/spacecrafts using In-Situ Mars Propellant Plant for Deep Space Transportation and earth departure for money.
- In forthcoming years, private agencies whoever wants to establish colony and be a member of Mars colony, they should pay some amount and get license to form colonies.
- Extracting and fabricating materials from Mars minerals and exporting them to earth for money.

**Administration**

Administration will be undertaken by the nations which are capable of fulfilling the technological needs of Mars colony or otherwise the administration will be alliance between nations. A nation will rule the colony for particular periods, then the next country, and then the next. It will be like a ruling cycle.

**Marship-I Assembly Order**

Thrusters-Fuel Tank-[Solar Panels] Cargo Module [Solar Panels]-(Crop Culture Module-I) Crewed Module (Crop Culture Module-I)-Life Support System-Medical Module-Ascent Vehicle-I

**Marship-II Assembly Order**

Thrusters-Fuel tank-[Solar Panels] Cargo Module [Solar Panels]-[Solar Panels] Lander Module [Solar Panels]-Rover Module-Habitat-I-Habitat-II-Ascent Vehicle-II

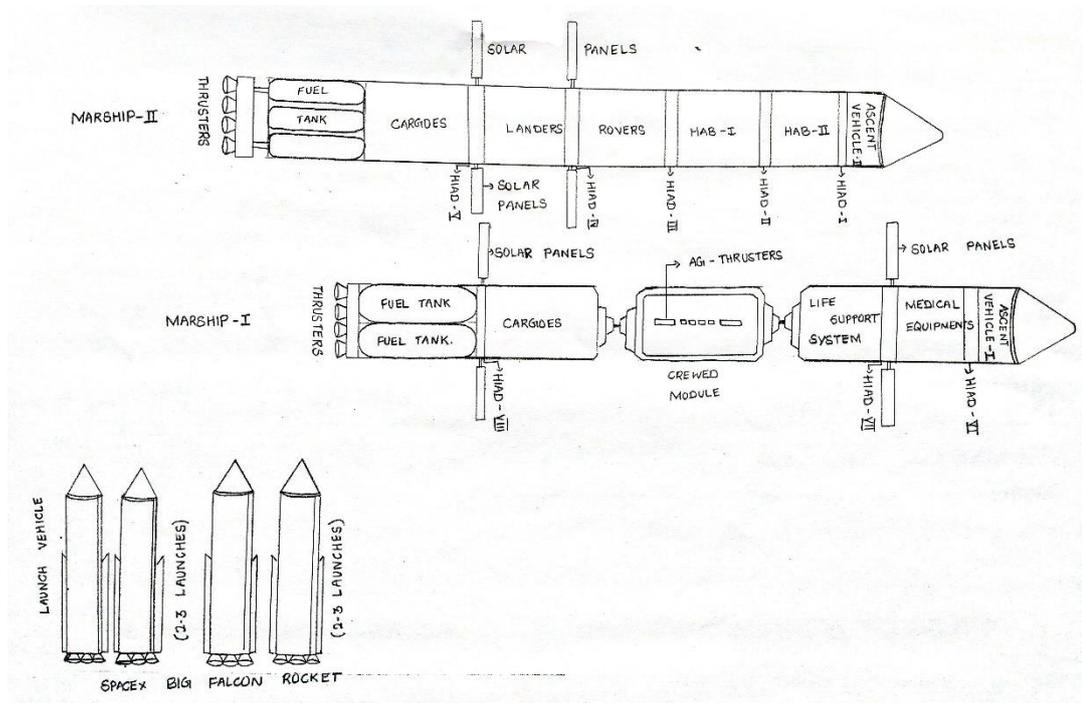

Fig.1. Phase I – Launch and Assembly of Marship I &II

**Thrusters:** Vacuum thrusters to propel the spaceship towards Mars is powered by either organic fuels or nuclear fusion reactor.

**Fuel Tank**: Fuel tank mostly consist of MMH (mono-methyl hydrazine or hydrazine) for spaceship propulsion. The tank is customizable and will be utilized for crop culture in future.

**Lander Modules**: Lander modules consists of scientific equipment's, scientific laboratories and other miscellaneous materials or tools.

**Rover Modules:** It consists of Water Extraction Rover for water extraction, Mobile Rovers for astronaut movements, Construction Rovers for carrying raw materials to construction sites, Scientific Rovers for scientific activities and resource location.

**Habitat-I & II:** It will provide living space for astronauts on the Martian surface.

**Ascent Vehicle I & II (Optional)**: They are optional in case of return trip to Earth. And it can be used for travelling and accessing International Mars Station in Mars orbit.

**Crewed Module**: Crewed module provide space for travelling peoples. Additionally, it also possess AG thrusters to simulate artificial gravity. This module has attached crop culture module in perpendicular direction for sunlight exposure.

**Life Support System Module**: It holds the place for artificial growing plants grown under the exposure of artificial light (solar powered). This module is the place for enhancement of $CO_2$ and $O_2$ Cycle between humans and plants.

**An Approach to Communication**: Communication relay between Earth and Mars is a major challenge in effectuating manned and unmanned mission to Mars. So, sending signals to Mars directly from Earth may not be an efficient method to control landers or rovers on Mars. It may cause serious issue during EDL guidance. Signals from earth takes an average time lag range from 13minutes and 48 seconds to 24 minutes and minimum time lag achieved is 8 minutes (Source: Curiosity Rover). Another issue is during the blackout periods occurs due to solar conjunctions for (approx...15days for every 26 months) may cause communication delay between the missions and ground (Earth). So, for carrying out effective human and landing missions, we suggest the following method.

- International Mars Station (Option-1)
- Mars Master Control Station (Option-2)

Both the station is capable of receiving and amplifying the control signals from earth as well as automating and auto-correcting the signals according to the missions (Here all the mission plans and their programs were programmed in advance)

- **International Mars Station (Option-1)**
    - Automated Operation / Manual Operation
    - Most Similar to International Space Station
    - Have Hospitality Services for Astronauts
    - Provide better place for desiring and planning
    - Provide easy access to Mars and Beyond
- **Mars Master Control Station (Option-2)**
    - Fully dedicated for Signal Processing/Signal Amplification
    - Dual Computer/Dual Communication System (in case of system failure)
    - Placing in either Mars orbit or Planting on Martian moons (Phobos or Demos)
    - Operation Mode - Fully automated

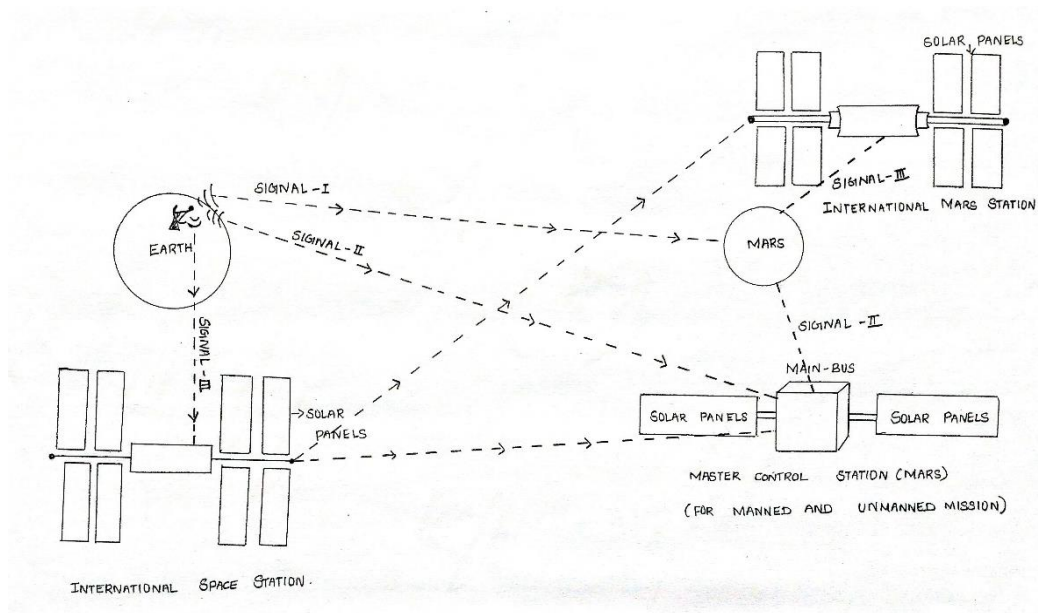

**Fig.2. An Approach to Communication System**

**Stage Separation & Preparation for Entry:** Figure depicts the separation of modules from Marship. Each module have attached HIAD for faster Aerocapture and will follow entry from Mars orbit.

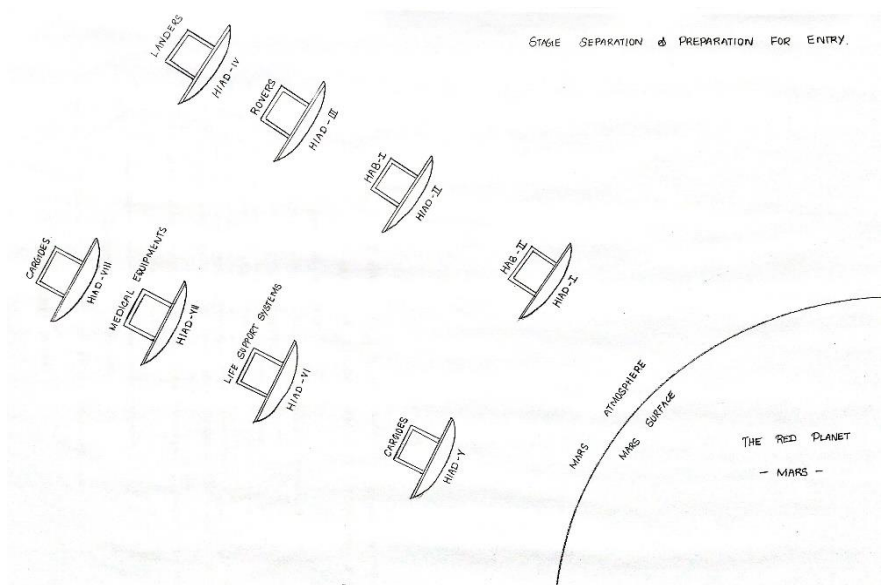

**Fig.3. Stage Separation and Preparation for entry**

**Crewed Module:** The crewed module comprising of restaurant, gymnastics, toilet, bath room, discussion room and power & maintenance room. A space specially allocated for sleeping called Sleeping Chamber that encloses Sleeping Berths (I, II & III). Similarly, a space allotted for scientific laboratories and doing scientific activities. Here A1 & A4 were primary air-locks and A2 & A3 were secondary air-locks, both the air-locks aids in maintaining air and pressure inside modules (also prevents from air escape). D1 & D2 were door locks to enter into the sleeping chamber.

- Restaurant – for feeding astronauts onboard Marship
- Discussion Room – for teaching astronauts about mission aspects
- Gymnastics – for making astronauts healthier
- Power & Maintenance Room – for maintaining power supply and mechanical maintenance.

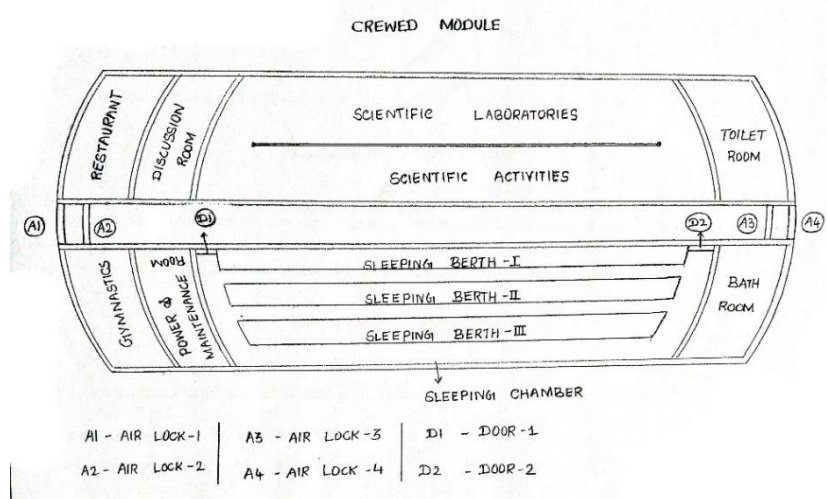

**Fig.4. Crewed Module**

**Crop Culture Module:** The upper portion of this module is covered with transparent glass in order to allow sunlight exposure to the growing plants. The lower portion of the module is allotted for plantation, soil maintenance and food production. Right corner of this module has a room for keeping farming tools and agricultural accessories. And another room allotted for Atmosphere and water regulation for plantation. Left corner of this module allocated for storing crops and foods. There is a power and maintenance room in every module. This module is aimed to rendezvous with crewed module in perpendicular direction to provide artificial gravity to growing plants. It also has primary and secondary air-lock systems.

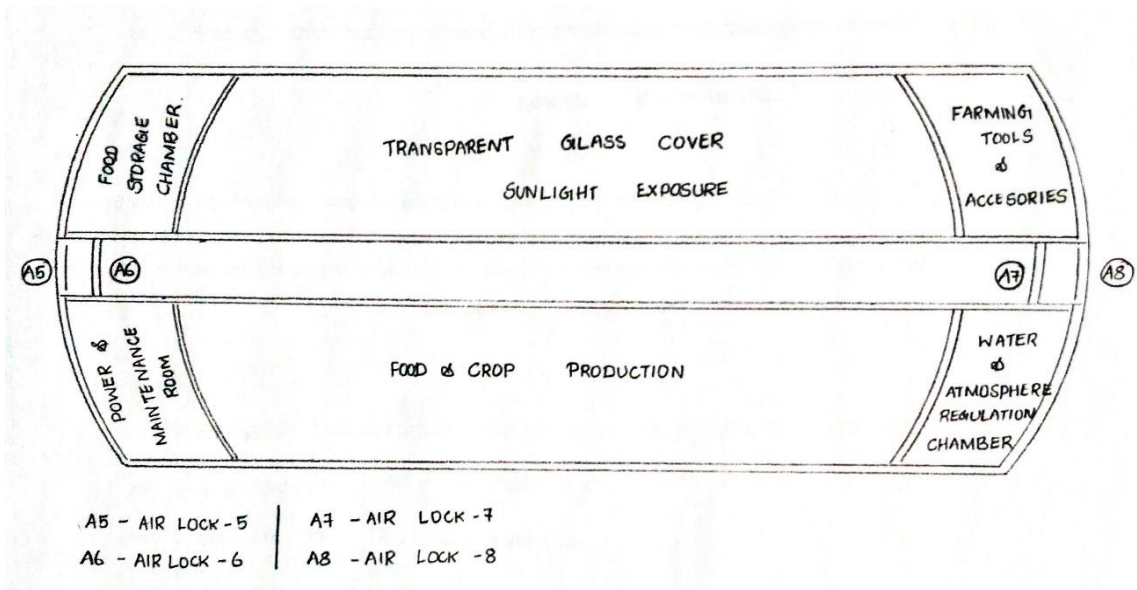

**Fig.5. Crop Culture Module**

**Medical Module:** Half-portion of this module is allotted for medical equipment and tools, similarly another half portion of this module is allotted for Deep Sleep of astronauts (i.e. inducing cryogenic sleep method for long distance travel). Right to this chamber, there is a room for cryogenic maintenance. Like all module it has power & maintenance room as well as primary and secondary air-locks.

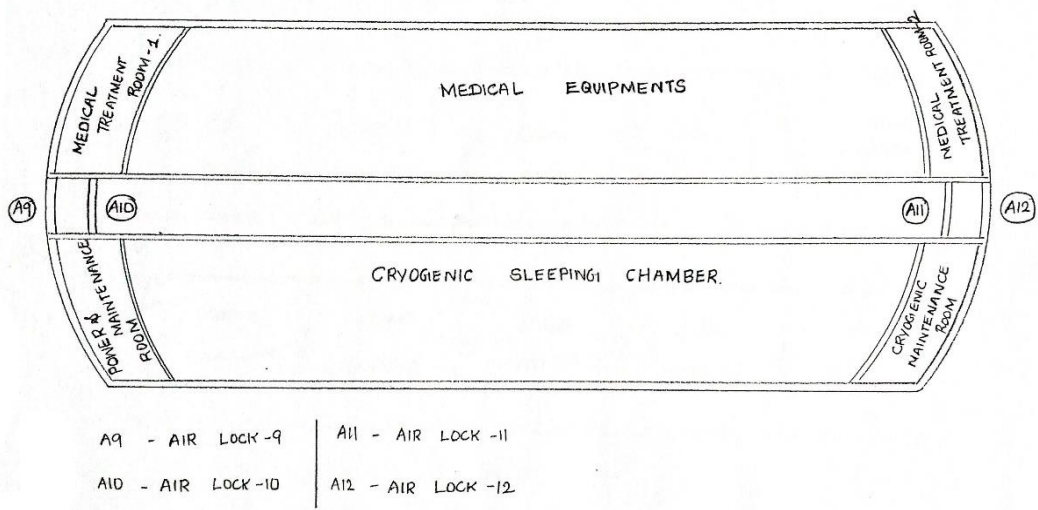

**Fig.6. Medical Module**

**Cargo Chamber/Module:** The whole cargo module, we divided into ten (10) cargo counters for loading and unloading cargoes, similarly we have made ten (10) ways to provide access to cargoes. It also possess refrigeration system and power & maintenance room, primary and secondary air-locks.

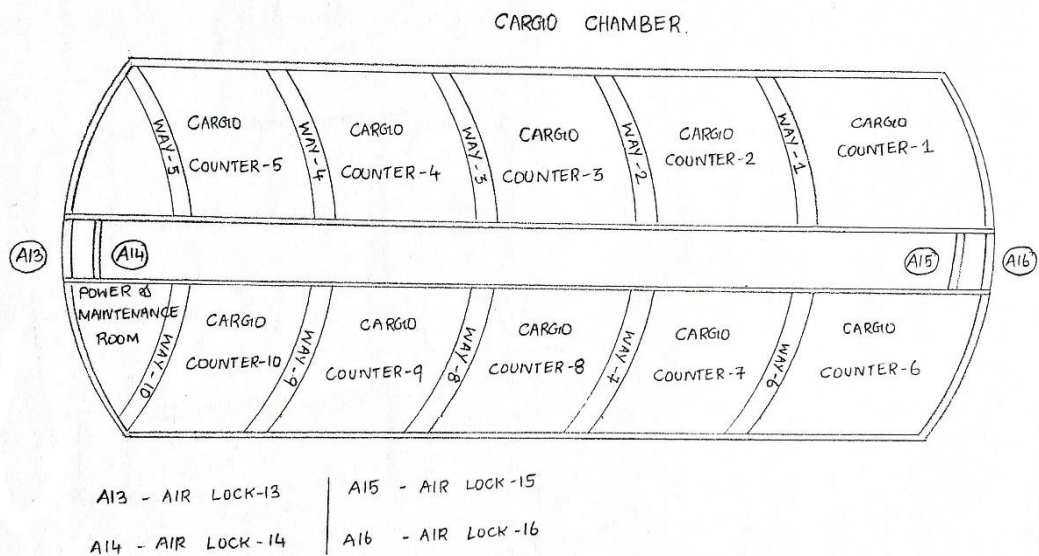

Fig.7. Cargo Module

**Mars Ascent Vehicle I & II:** They are additionally added (optional). In case of return trip to earth, Mars Ascent Vehicle will be used. Its major feature is delivering large masses to Earth and Mars orbit.

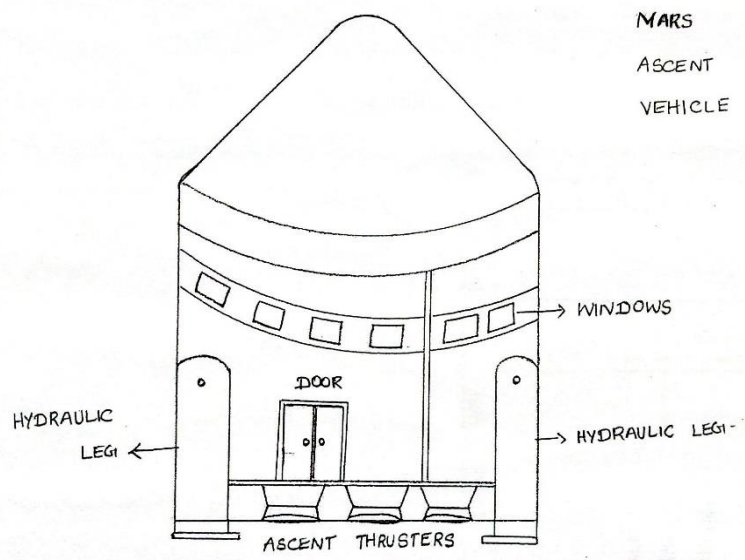

Fig.8. Mars Ascent Vehicle

**Habitat I & II:** The image depicts deployable solar panels attached to the hab and all modules. It is expandable and dust proof due to the dual slider cover technique. This technique may prevent the solar panels form covering dust encountered during dust storms.

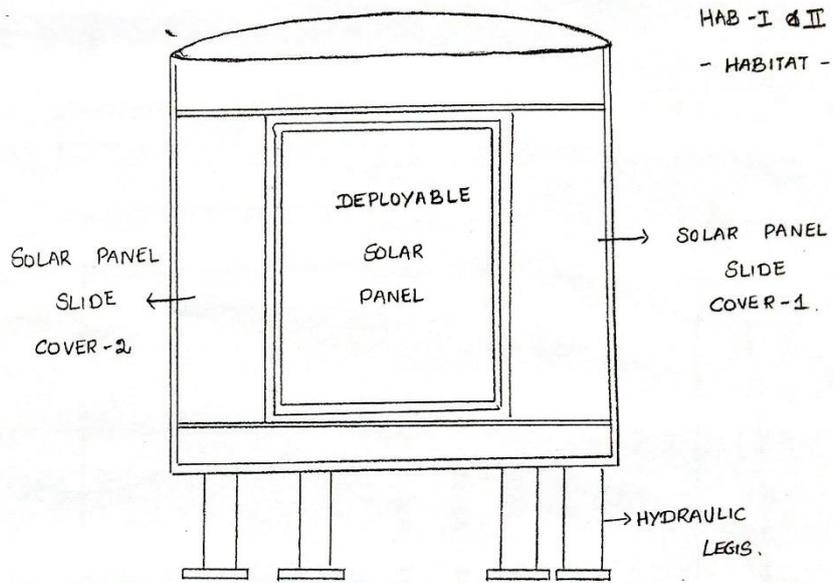

**Fig.9. Mars Habitat with deployable Solar Panels**

**Mars Sub-Surface Habitat:** One of the special feature of this habitat is to protect astronauts/laboratories/habitats from Martian robust dust storms [4]. The bottom portion of this hab has sub-surface miner and side portion has screw like heads. The whole hab will penetrate into the surface. Inner portion of this hab has two hydraulic hab lifters that will lift the hab to the surface during daytime for research activity or for sunlight exposure. And during night time (in order to maintain thermal stability) or during devil dust storms the hab will be lifted down into the sub-surface for protection.

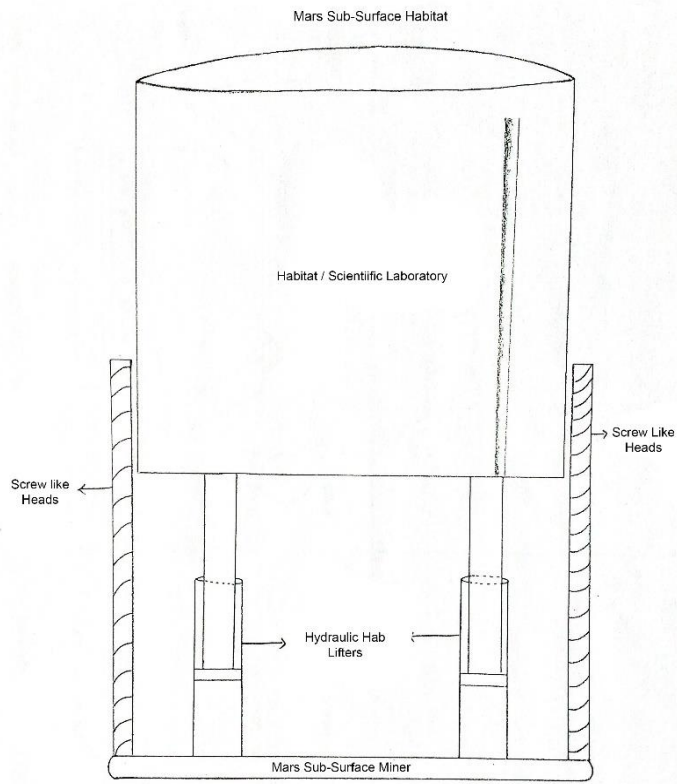

**Fig.10. Mars Sub-Surface Habitat**

**Manual Sub-surface Ice Drillers:** In addition to "Mars Sub-surface Water Extraction Rover", manual sub-surface ice drillers will be used manually by astronauts to extract water from the sub-surface ice.

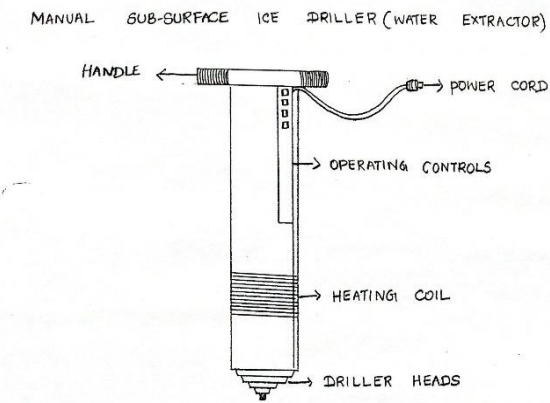

**Fig.11. Manual Sub-Surface Ice Driller**

**Water Extraction Rover:** Due to the presence of sub-surface ice [5]. We can extract water. So water extraction rover possess driller heads at the bottom part and four supporting arms. The arms are capable of moving the rover body up and down in a convenient way to extract water from surface.

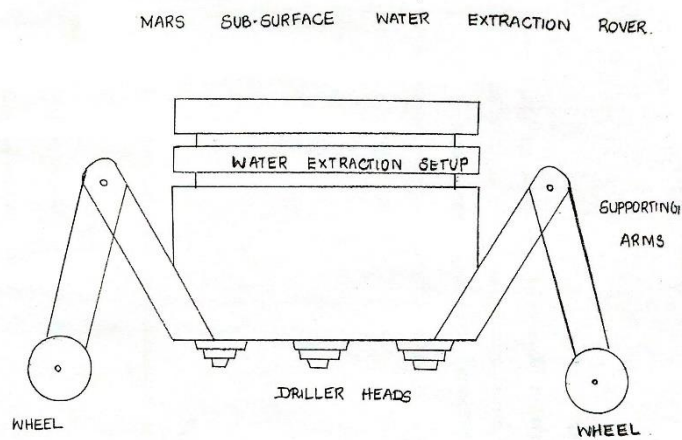

**Fig.12. Mars Sub-Surface Water Extraction Rover**

**Crop Culture System:** As mentioned earlier, customizable fuel tanks again re-customized and engineered on the surface for crop culture (optional). A new crop culture setup will be carried out from Earth. It has upper block for fitting artificial robotic arms for crop maintenance and the lower block for soil and plantation. It also possesses mechanical gravity simulator to simulate artificial gravity and three ports for passage and exchange of $CO_2$, $O_2$ and $H_2O$.

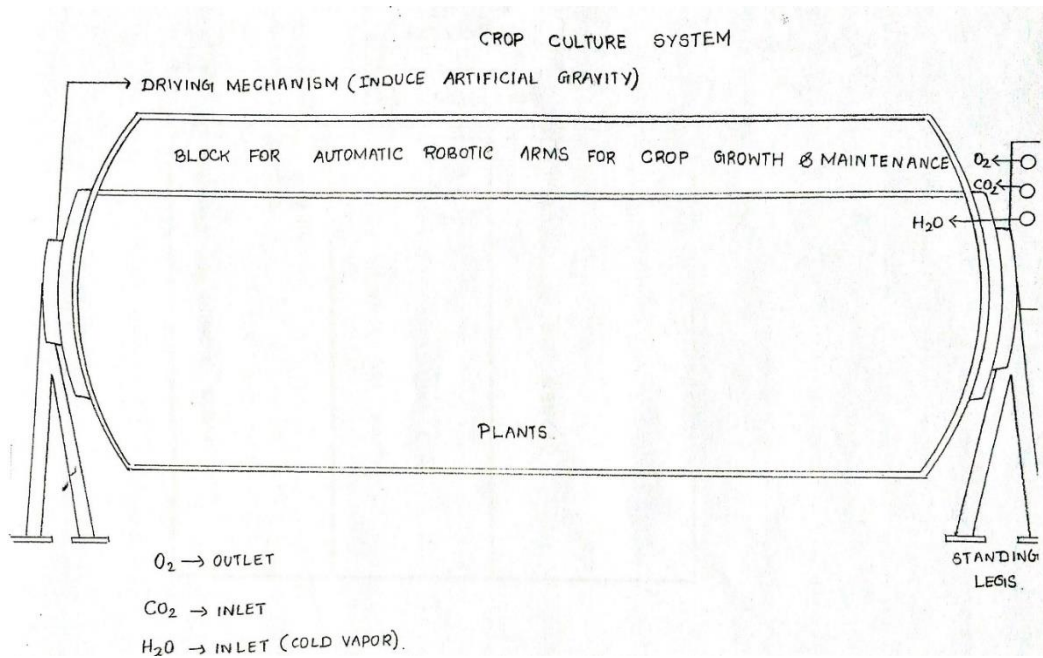

**Fig.13. Crop Culture System**

**Atmosphere Regulation System:** Atmosphere regulation system possesses three tanks oxygen, carbon dioxide and water tanks. Top portion this tank has $CO_2$, $O_2$ and $H_2O$ regulator. This regulator has a port to absorb atmospheric $CO_2$, a port to absorb water from Rover, a port for giving out oxygen for astronauts.

Similarly, $CO_2$ tank has $CO_2$ port to supply $CO_2$ to Crop culture system, $H_2O$ tank has $H_2O$ port to supply water in the form of moist air and $O_2$ tank has $O_2$ port to absorb $O_2$ released by plants inside crop culture system in order to produce oxygen for astronauts.

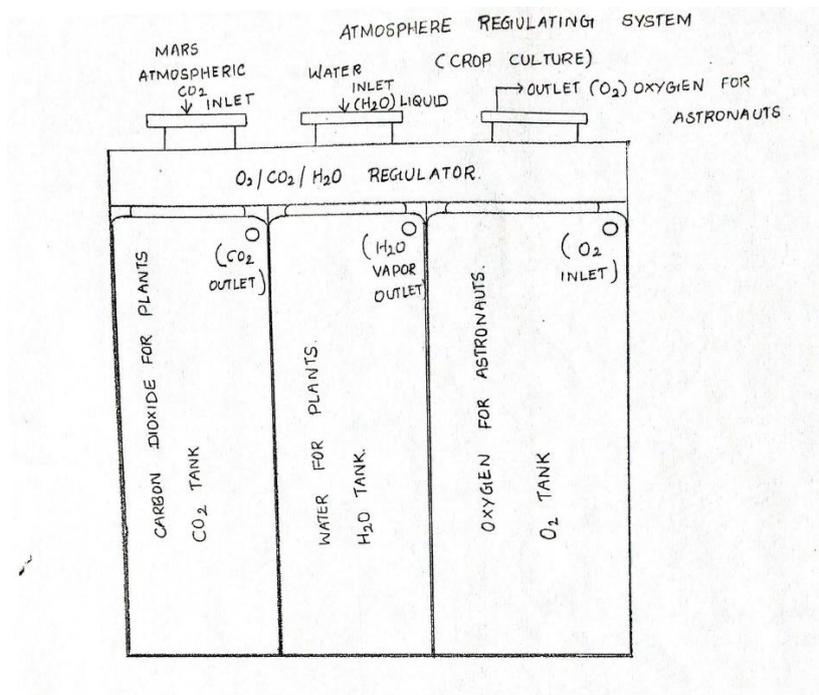

**Fig.14. Atmosphere Regulating System**

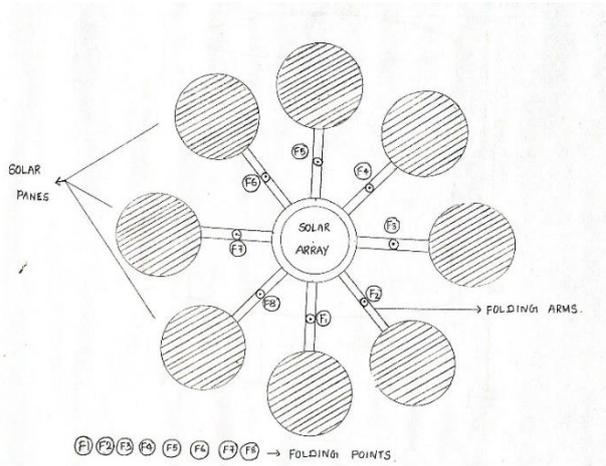

**Mars Solar Power Plant:** It comprises of 8 Solar Panels which are enclosed inside Solar Array through foldable arms. Using these type of Solar Array, we can enclose more number of solar panels inside a single module.

**Fig.15. Mars Solar Power Plant**

**Solar Cell Radiation Shield:** Here we are trying to explain that, solar cells should be made using some radiation shield metals and that solar panels can be placed over the habitat. As a result the solar panel will protect the astronauts from harmful radiation.

**Fig.16. Solar Cell Radiation Shield**

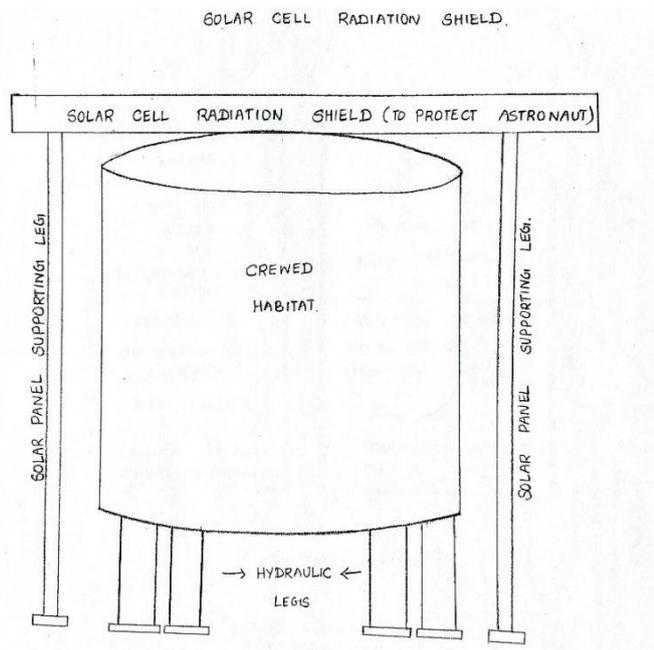

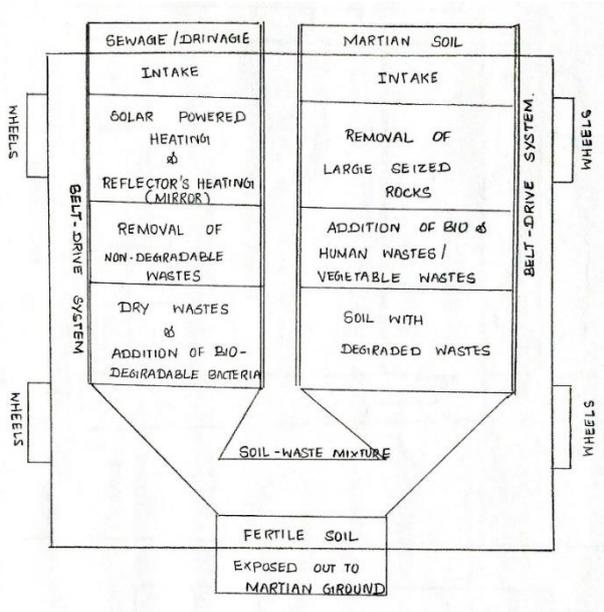

**Mars Soli Fertilizing Rover:** This is a rover which will be converting the un-fertilized Martian soil into fertilized type using the waste generated by crews and vegetable wastes. This rover will choose the location and will map the area in square meter then it will move over the soil as per the map given below.(1-9 Moving Tracks)

**Fig.17. Mars Soil fertility Rover / Map**

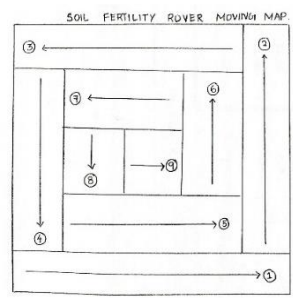

**Rock Sizing Plant:** This picture depicts the plant which sizes the rock and slices into pieces. The rocks which are comes as output is loaded to the Mars Load Rover, and then it get transported to the construction site.

**Iron Production Plant:** This plant produces iron from the ferrous ore that are mixed with the Martian soil through smelting process.

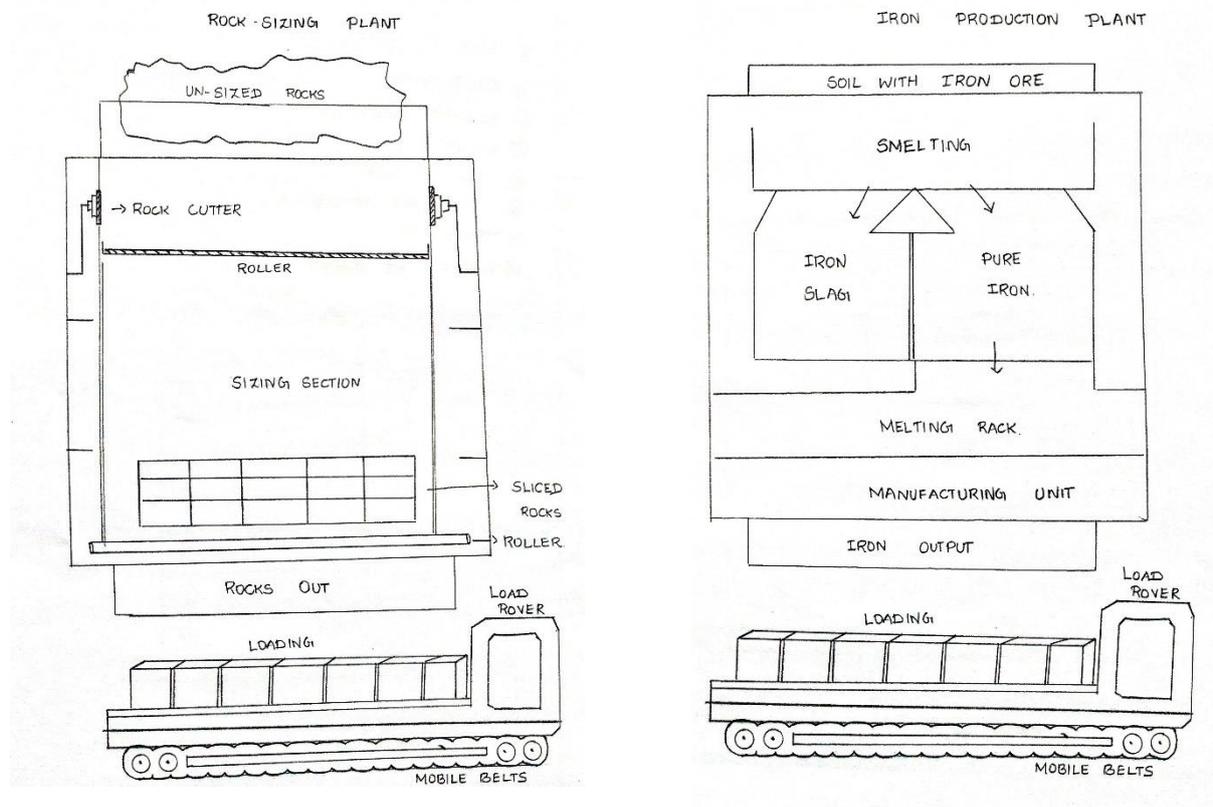

Fig.18. Mars Rock Sizing & Iron Production Plant

**Mars Propellant Plant:** This plant produces fuel for propulsion systems using Mars atmospheric air ($CO_2$). It electrolysis water into hydrogen and oxygen. Then it combines with the carbon dioxide to form methanol.

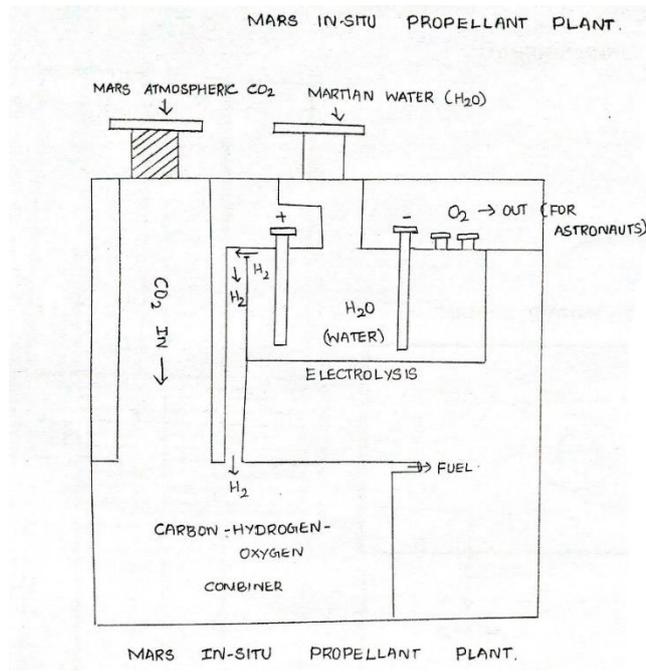

**Fig.19. Mars In-Situ Propellant Plant**

**Water Extraction Setup from Rover:** The setup has 8 layers to purify water.

1. Drilled Ice Stacks
2. Heating Coil
3. Water from melted ice
4. Martian Water
5. Filter bed
6. Purified Water
7. Chlorination
8. Pure Martian Water (Aqua Mars)

The driller heads at the bottom of the setup, drills and extracts ice from the sub-surface of Mars. Then a heating coil powered by solar heats and melts the ice into impure water. After this process, the water passes through the filter bed where the muddy water get purified into clean water. Following this stage, the water passes through the chlorination rack where water becomes completely pure that can be used as drinking water for astronauts as well as planting crops. (Aqua Mars).

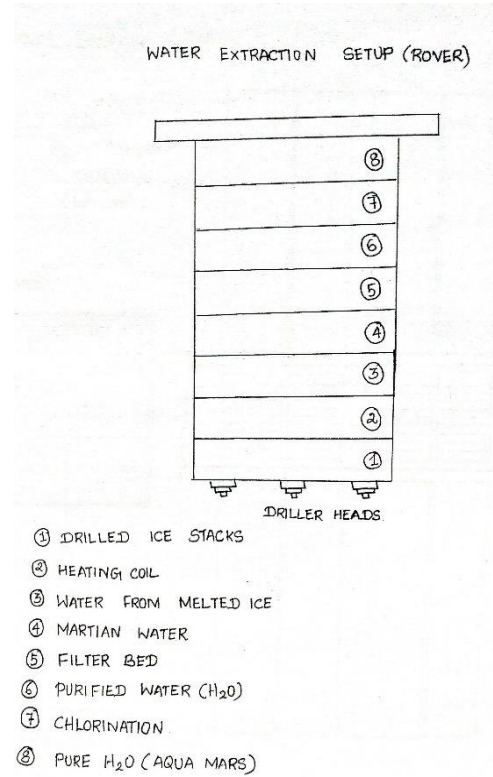

**Fig.20. Water Extraction Setup from Rover**

**Mars Load Rover:** Mars load rover will be used to carry loads from one place to another. The rover is solar powered and it has conveyor belt. This provides better movement in sandy regions like Mars surface. The rover is automated as well as manually operated by astronauts. It has flash and night lamps for the use in night time.

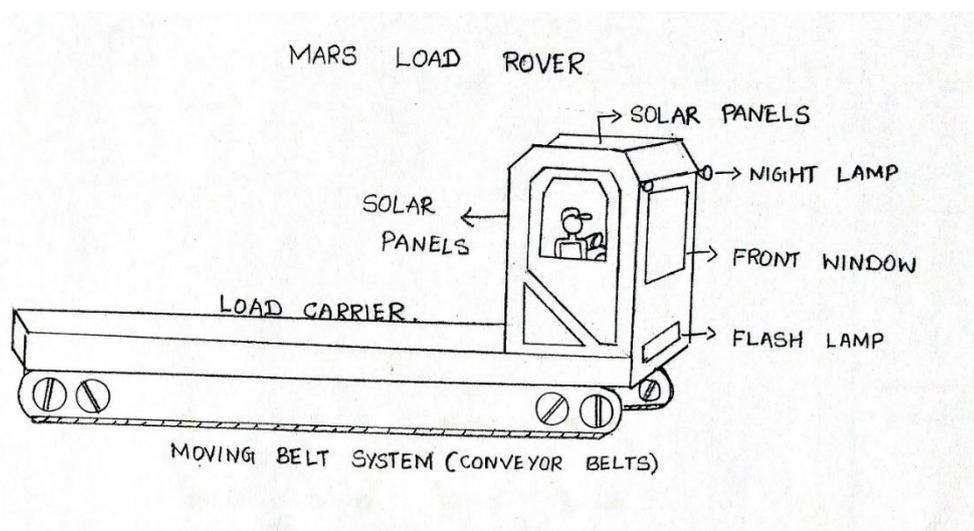

**Fig.21. Mars Load Rover**

**Mars Passenger Rover:** Mars Passenger Rover, it will allow astronauts to move in groups over Martian surface. This rover is also similar to load rover. Solar powered and conveyor belt moving system. Automated as well as manually operated.

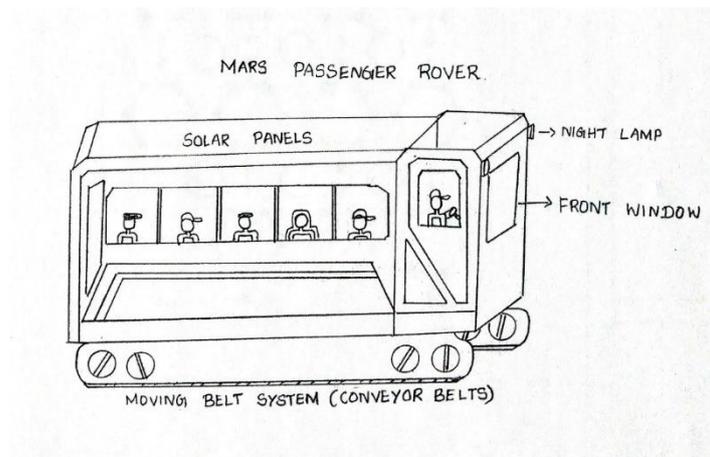

**Fig.22. Mars Passenger Rover**

**International Mars Station:**

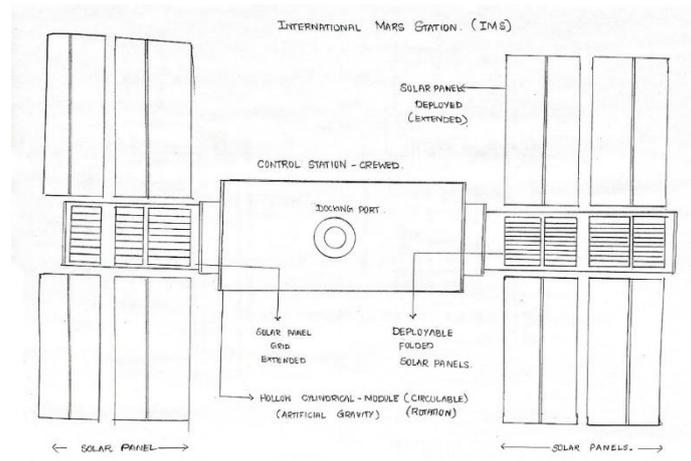

**Mars Minerals and their Applications: [6][7-12]**

| Minerals | Applications |
| --- | --- |
| Olivine | Refractory Materials |
| Marcassite | Diamond and Graphite |
| Feldspar | Dinnerware, Bathroom and building tiles, Glass production, Flux |
| Ilmenite | Aircraft parts, Titanium metal, Reflective pigments, Synthetic rutile |
| Akaganeite | Adsorption of arsenic to prevent pollution |
| Hematite | Storage Devices, Electrical components, Inorganic pigments |
| Magnetite | Water purification, Biomedical and environmental applications |
| Goethite | Cleaning chemical elements in polluted water |
| Saponite | Water Decontamination, Rubber, Cosmetics and Medicine |
| Montmorillonite | Nuclear applications, Cotton fiber, Adsorption in heavy metals and dyes |
| Serpentite | Architecture, Ornaments, Asbestos |
| Illite | Ceramics, Tiles |
| Kieserite | Cleaning hard water, Epsom salt, fertilizer |
| Gypsum | Cement, Plaster of Paris |

**Calculation of food consumption and mass transportation:[13]**

| S.No | Item | Approx. Food Consumption | Consumption /day/person | For 1000 people/day | For 1000 peoples/300 days | Extra for usage on Mars |
|---|---|---|---|---|---|---|
| 1. | Food | 2 kg | 1.8 kg | 1800 kg | 540,000 kg | 180,000 kg |
| 2. | Water | 4.17 kg | 5 kg | 5000 kg | 1,500,000 kg | 500,000 kg |
| 3. | $CO_2$ | 1 kg | 1 kg | 1000 kg | 300,000 kg | 100,000 kg |
| | | | | Total | 2,340,000 kg | 780,000 kg |
| Number of Big Falcon Rocket Required for launches = 17 – 18 BFR's ||||||||
| Cost estimated for this launch = $90 Billion USD Dollars ||||||||

**Suggestion for faster food and oxygen production using plants:**

| Food Production [14] | Research Plants [15] | Ornamental Plants [15] | Oxygen Producing [16] |
|---|---|---|---|
| Arugula | Arabidopsis thaliana | Zinna | Areca Palm |
| Spinach | Brachypodium distachyon | | Snake Plant |
| Carrots | Brassica rapa | | Money Plant |
| Cucumbers | Ceratopteris richardii | | Gerbera Daisy |
| Beet Roots | Mizuna Lettuce | | Chinese Evergreens |
| Bush Beans | Zucchini | | |
| Bok Choy | Wheat | | |
| Lettuce | Red Ramine Lettuce | | |
| Summer Squash | Rice | | |
| Okra | Tomato | | |
| Kale/Greens | Spinach | | |
| Snow Peas | Pepper | | |
| Broccoli | Broccoli | | |
| Green Onions | | | |
| Turnips | | | |
| Radishes | | | |

**Activities | Arts | Sports**

- **Martian School Activity:** In schools, the following will be taught - basic English, physics and it's all branches will be taught in accordance to Martian gravity, temperature, How to do space agriculture?, How to avoid space accidents, how to escape, how to survive in a place using available resources, how to manage space activities, briefing and explaining about space science, explaining our past and where we came from, making them to realize the beauty of life, how to maintain discipline in space travel, how to manage our bases.
- **Arts Activity:** Creating new music's, dancing, visual screening of life on earth, promoting and encouraging handicrafts making using available resources.
- **Sports Activity:** Running and Jogging, Gravity related games, jumping games, breathing games and survival games that will be in favor of Mars survival.

## Conclusion:

With reference to the guidelines of Mars Society technical design of human Mars mission architecture were proposed, the beauties of colony can be made by establishing and constructing bases according to the pattern shown in second page "Human Mars Settlement Map" activities were shortlisted, administration will be based on united alliance between nations, economic aspects of Mars colony is explained, suggestions and considerations for carrying out human Mars missions are clearly defined.

## Nomenclature:

| | | |
|---|---|---|
| IMS | - | International Mars Station |
| HAB | - | Habitat |
| MMCS | - | Mars Master Control Station |
| MAV | - | Mars Ascent Vehicle |
| HIAD | - | Hypersonic Inflatable Aerodynamic Decelerator |
| ISS | - | International Space Station |
| AG | - | Artificial Gravity |
| BFR | - | Big Falcon Rocket |
| LP | - | Landing Pad |

## Acknowledgement

The authors would like to thank Dr.Robert Zubrin and all the members of Mars Society for providing us opportunity to participate in designing plans for Human Mars Settlement Contest.